\documentclass[journal]{IEEEtran}
\usepackage[T1]{fontenc}

\usepackage{tikz}
\usetikzlibrary{spy}
\usepackage{algorithm}
\usepackage{algorithmicx}
\usepackage{algpseudocode}
\usepackage{amsfonts}
\usepackage{amsmath}
\usepackage{amssymb}
\usepackage[utf8]{inputenc}
\usepackage{xcolor}
\usepackage{mathtools}
\usepackage{graphicx}
\usepackage{pgfplots}
\pgfplotsset{compat=newest}
\usetikzlibrary{plotmarks}
\usetikzlibrary{arrows.meta}
\usepgfplotslibrary{patchplots}
\usepackage{grffile}

   \pgfplotsset{plot coordinates/math parser=false}
   \newlength\fheight
   \newlength\fwidth
\usepackage{caption}
\usepackage{subcaption}
\usepackage{import}
\usepackage{threeparttable}
\usepackage{tabularx}
   \usepackage{multirow}
\usepackage{booktabs}
\usepackage{cite}
\usepackage[export]{adjustbox}
\usepackage{breqn}
\usepackage{mathrsfs}
\usepackage{acronym}
\usepackage[acronym]{glossaries}
\usepackage{setspace}
\usepackage{empheq}
\usetikzlibrary{patterns}

\DeclareMathOperator*{\argmax}{arg\,max}

\usepackage[font=footnotesize]{subcaption}
\usepackage[font=footnotesize]{caption}

\graphicspath{{./}}
\newcommand{\fig}[1]{Fig.~\ref{#1}}

\newcommand{\secref}[1]{Section~\ref{#1}}

\newacronym{dsrc}{DSRC}{Dedicated Short Range Communication}
\newacronym{mmwave}{mmWave}{millimeter wave}
\newacronym{lte}{LTE}{Long Term Evolution}
\newacronym{v2i}{V2I}{Vehicle-to-Infrastructure}
\newacronym{v2n}{V2N}{Vehicle-to-Network}
\newacronym{v2v}{V2V}{Vehicle-to-Vehicle}
\newacronym{ue}{UE}{User Equipment}
\newacronym{wave}{WAVE}{Wireless Access in Vehicular Environments}
\newacronym{wimax}{WiMAX}{Worldwide Interoperability for Microwave Access}
\newacronym{los}{LOS}{Line of Sight}
\newacronym{nlos}{NLOS}{Non Line of Sight}
\newacronym{ber}{BER}{Bit Error Rate}
\newacronym{snr}{SNR}{Signal-to-Noise Ratio}
\newacronym{bs}{BS}{base station}
\newacronym{ci}{CI}{Confidence Intervals}
\newacronym{gnb}{gNB}{Next Generation Node Bases}
\newacronym{enodeb}{eNodeB}{4G LTE base stations}
\newacronym{nr}{NR}{New Radio}
\newacronym{iab}{IAB}{Integrated Access and Backhaul}
\newacronym{si}{SI}{Study Item}
\newacronym{rsu}{RSU}{Road Side Unit}
\newacronym{iot}{IoT}{Internet of Things}
\newacronym{3gpp}{3GPP}{Third Generation Partnership Project}
\newacronym{isd}{ISD}{Inter-site distance}
\newacronym{hetnet}{HetNet}{heterogeneous network}
\newacronym{rat}{RAT}{Radio Access Technology}
\newacronym{rss}{RSS}{Received signal strength}
\newacronym{maxRSS}{max-RSS}{maximum received signal strength}
\newacronym{qos}{QoS}{Quality of Service}
\newacronym{c-its}{C-ITS}{Cooperative and Intelligent Transportation System}
\newacronym{lidar}{LIDAR}{Laser Imaging Detection and Ranging}
\newacronym{mimo}{MIMO}{Multiple Input Multiple Output}
\newacronym{h2h}{H2H}{human-to-human}
\newacronym{m2m}{M2M}{machine-to-machine}
\newacronym{vn}{VN}{vehicular node}
\newacronym{ms}{MS}{maximum SNR}
\newacronym{mr}{MR}{maximum rate}
\newacronym{ra}{RA}{requirement-aware}
\newacronym{ppp}{PPP}{Poisson Point Process}
\usepackage{bbm}

\def \eqPLLTE{
\begin{equation}
\mathbb{P}_{\rm LOS}^{\rm LTE}(d) = \min \Big( \tfrac{0.018}{d},1 \Big) \left[ 1-\exp\left(\frac{-d}{0.063}\right)\right]+
\exp\left(\frac{-d}{0.063}\right),
\end{equation}
}

\def \psat{
\begin{equation}\label{eq:perc_satisfied}
  p_{{\rm sat}} = \displaystyle \frac{\sum_{i \in \mathcal{M}_k} \mathbf{1} [R_{ij^*(i)}>\bar{R_i}]}{|\mathcal{M}_k|} \quad \forall k\in\{1,\dots,4\}
  \end{equation}
}

\def \eqPLmmw{
\begin{equation}\label{eq:pl_mmw}
\mathbb{P}_{\rm LOS}^{\rm mmW}(d)=
  \begin{cases}
    1 & \text{for} ~d \leq 18 \text{ m}\\
  \frac{18}{d} + \exp\left(-\frac{d}{36}\right)\left(1 - \frac{18}{d}\right) & \text{for} ~d > 18\text{ m}
  \end{cases}
\end{equation}
}

\title{An Efficient Requirement-Aware Attachment Policy for Future Millimeter Wave Vehicular Networks}

\author{\IEEEauthorblockN{ Davide Peron, Marco Giordani, Michele Zorzi\\}
\IEEEauthorblockA{
Department of Information Engineering (DEI), University of Padova, Italy \\
Email:{\texttt{\{perondav,giordani,zorzi\}@dei.unipd.it}}}}

\IEEEoverridecommandlockouts
\begin{document}

\maketitle
\thispagestyle{empty}
\pagestyle{empty}

\begin{abstract}

The automotive industry is rapidly evolving towards connected and autonomous vehicles, whose ever more stringent  data traffic requirements might exceed the capacity of traditional technologies for vehicular networks.  In this scenario, densely deploying millimeter wave (mmWave) base stations is a promising approach to provide very high transmission speeds to the vehicles. However, mmWave signals suffer from high path and penetration losses which might render the communication unreliable and discontinuous.
Coexistence between  mmWave  and Long Term Evolution (LTE) communication systems has therefore been considered to guarantee increased  capacity and robustness through heterogeneous networking.
Following this rationale,  we face the challenge of designing fair and efficient attachment policies in heterogeneous vehicular networks. 
Traditional methods based on received signal quality criteria lack consideration of the vehicle's individual requirements and traffic  demands, and  lead to suboptimal resource allocation across the network.
In this paper we propose a Quality-of-Service (QoS) aware attachment scheme which biases the cell selection as a function of the vehicular service requirements, preventing the overload of transmission links.
Our simulations demonstrate that the proposed strategy significantly improves the percentage of vehicles satisfying application requirements and delivers efficient and fair association compared to state-of-the-art~schemes.

\end{abstract}

\section{Introduction}
\label{sec:introduction}

In recent years, there has been a significant interest in the context of \glspl{c-its}, which have rapidly emerged as a means to guarantee a safer travel experience and  to support multimedia services~\cite{bengler2014three}. 
The potential of   autonomous vehicles can be fully unleashed through direct wireless communications to and from roadside infrastructures, a concept that is usually referred to as \gls{v2i} networking.
Although the \gls{lte} standard presently represents the principal wireless technology offering \gls{v2i} transmission services~\cite{araniti2013lte}, future vehicular networks will 
have ever more stringent regulations in terms of road safety and traffic management~\cite{3gpp.22.186}. 
In particular, next-generation vehicles will be equipped with sophisticated instrumentation (e.g., high-resolution LIDAR and camera sensors) which will be the source of unprecedentedly high data rates (in the order of hundreds of gigabits per second, according to some estimates~\cite{lu2014connected}): LTE-based vehicular networks were primarily designed to provide coverage and are therefore unsuitable to accommodate such huge change in the traffic~trends. 

In this context, the automotive industry  has  devoted efforts to specifying new communication solutions, e.g., operating in the \gls{mmwave} bands above 10 GHz, that allow vehicles to use very large bandwidths to communicate, thus guaranteeing very high transmission speeds~\cite{choi2016millimeter}. 
Although this new band has gathered great attention for \gls{v2i} applications,  the \gls{mmwave} paradigm comes with its own set of challenges~\cite{giordani2017millimeter}. 
In particular,  \gls{mmwave} transmissions suffer from severe path loss and susceptibility to blockage, and prevent long-lived communications, a critical prerequisite for  safety~operations.

One promising approach to handle \gls{mmwave} limitations is to increase the density of the \glspl{bs}, to reduce inter-site distance and establish stronger access channels.
Massive \gls{mimo} techniques have also emerged as a means to provide an additional beamforming gain to the link budget, compensating for the increased propagation loss. 
Moreover, heterogeneous networking~\cite{zheng2015heterogeneous,Dhillon13} has recently been considered to improve network capacity, e.g., by combining a reliable sub-6 GHz link (e.g., using \gls{lte}) with a high-capacity mmWave connection.
Despite some encouraging features, heterogeneous networking deployments lack  at least one important design aspect, which is how to efficiently and fairly associate vehicles to the network.
Maximum downlink received power based association, for example, typically leads to a limited number of nodes actually getting served by \gls{mmwave} cells due to their much more unstable propagation characteristics  compared to \gls{lte} cells. 
Maximum rate based association, on the other hand, tends to prioritize \gls{mmwave} \glspl{bs} over legacy ones due to the much larger bandwidth available to high-frequency systems.
This load disparity inevitably leads to suboptimal resource allocation, with a large number of vehicles experiencing poor date rates in overloaded cells while the resources in other lightly loaded cells can be underutilized.

Following this rationale, in this paper we address the issue of balancing network association requests between \gls{lte} and \gls{mmwave} \glspl{bs},  avoiding the overload of transmission links.
To do so, we design a novel \gls{qos} aware attachment strategy that identifies the most appropriate destination cell as a function of the vehicle's individual requirements and traffic  demands.
Our results show that the proposed approach, which biases the received signal quality criteria with additional information about network loading and vehicle requirements, can significantly improve the connectivity performance compared to conventional association~solutions, even considering scarcely deployed networks.
It also offers good connectivity to cell-edge vehicles, i.e., the more channel-constrained network entities, and guarantees a fair distribution of the available resources across the cell. 

The rest of this work is organized as follows. In \secref{sec:related_work}, we discuss related work on heterogeneous networking and user association techniques, while in \secref{sec:attachment} we present the different attachment strategies we consider in our analysis.
In \secref{sec:performance_evaluation} we describe our simulation scenario and parameters, and present our main results and discoveries.
Finally, \secref{sec:conclusions} concludes the paper and gives suggestions on possible extensions for future research.

\section{Related Work}
\label{sec:related_work}

One of the main challenges of multi-radio heterogeneous networking is the design of the optimal user association technique that avoids system overload  and  best distributes the available resources among the users.




Different studies have been conducted in this field, the most relevant ones taking network metrics into account, e.g., the distance from the \glspl{bs}, path loss, \gls{snr} or data rate.
For example, Chen \emph{et al.}, in~\cite{chen2011joint}, proposed joint optimization of channel
selection, user association and power control to maximize spectrum and energy
utilization efficiencies.
In~\cite{corroy2012dynamic}, Corroy \emph{et al.} presented a new theoretical framework to study cell
association for the downlink of multi-cell networks and developed a
dynamic method that  associates users to macro or pico nodes  while maximizing the sum rate of all network users. 
Cell range expansion theory has also been used in \cite{jo2012heterogeneous} to perform user association
based on the biased measured signal, i.e., balancing the load
among high- and low-power \glspl{bs}.

Lately, researchers have tried to solve the user association problem using advanced mathematical tools, in particular game theory~\cite{lasaulce2011game} and combinatorial optimization~\cite{papadimitriou1998combinatorial}.
For instance, in~\cite{bethanabhotla2014user}, the authors have  proposed load balancing  methods  for multi-tier networks with massive \gls{mimo} \glspl{bs} and demonstrated that the load-based association scheme  terminates in a Nash equilibrium.
Similarly, Xu \emph{et al.}, in~\cite{xu2017user}, presented a  centralized
user association algorithm that targets rate maximization, proportional fairness,
and joint user association and resource allocation in a \gls{mimo} scenario.
In~\cite{Liu14}, game theory was used to model user association in heterogeneous networks to guarantee \gls{qos} to human-initiated traffic while providing fair resource allocation for machine-to-machine services.
In \cite{Liu16}, Liu \emph{et al.}  formulated the user
association issue as a nonlinear combinatorial problem and proposed a centralized  scheme which guarantees fair and energy efficient attachment through Lagrange multipliers.
In~\cite{ye2013user}, the authors formulated a logarithmic utility maximization problem for
single-BS association, and showed that equal resource allocation
is actually optimal, over a sufficiently large time window.
However, most popular mathematical optimizations  only apply to scenarios where the traffic flow generated by endnodes is approximately static. However, in the real world, traffic is not stable nor accurately predictable, thereby making traditional model assumptions invalid.

Stochastic geometry~\cite{baccelli2010stochastic} has also emerged  as a computationally tractable approach to model and analyze the performance of multi-tier heterogeneous networks~\cite{elsawy2013stochastic}.  
In this regard, Dhillon \emph{et al.}, in~\cite{Dhillon13}, exploited stochastic geometry to evaluate the performance of  user association, based on received signal quality criteria, in a multi-tier cellular system.
In a similar way,  the authors in~\cite{Cheung12}  formulated a throughput maximization
problem subject to \gls{qos} constraints, and provided
insights into the optimal spectrum allocation technique.

Most prior work on network association  applies to
 LTE-only scenarios. \gls{lte} and \gls{mmwave}  heterogeneous networking, on the other hand, is much more sensitive to
the cell association policy because of the significant propagation disparities
of the two radios, and  calls for innovative solutions that depend on the radio technology characteristics.
In \cite{Singh2015}, Singh \emph{et al.} made the case that, although mmWaves generally represent the preferred access technology,  offloading users  to  more reliable radio interfaces may dramatically improve the rate of cell-edge users in case of sudden channel degradation.
Similarly, the authors in \cite{giordani2018efficient} proposed a novel uplink measurement system that, with the joint help of a local administrator operating in the \gls{lte} band, coordinates  user association requests as a function of the instantaneous load conditions of the surrounding cells, thereby promoting fairness in the network.

The aforementioned association policies were proposed for cellular networks, which might not be fully representative of a vehicular system due to the  more challenging propagation and traffic characteristics of highly mobile vehicular nodes.
Although some recent works in the literature have tried to provide preliminary insights into user association also in the context of vehicular networks~\cite{liang2018towards}, e.g., leveraging reinforcement
learning~\cite{li2017user}, there remain many open problems which call for innovative modeling and design solutions. 
Our work tries to fill this gap by extending traditional cellular-based attachment algorithms and  integrating physical-layer metrics with application requirements at the higher layers.

\section{Attachment policies}
\label{sec:attachment}

When a \gls{vn} enters a vehicular network for the first time, it needs to establish an initial physical link connection with a cell,  a procedure that is usually referred to as network attachment~\cite{giordani2016initial}. 
Traditional attachment procedures  monitor the quality of the received signals, which is typically expressed in terms of \gls{snr}, and select, as a target cell, the \gls{bs} from which the  maximum \gls{snr} was experienced.  This procedure is described in Sec.~\ref{ssec:max_SNR} and represents the benchmark solution of our analysis.
In this work we target tight integration of classic physical-layer performance metrics with additional network information in the upper layers. 
In particular, a maximum rate attachment policy, which takes data rate estimates into account, and a requirement-aware attachment policy, which biases cell selection as a function of the \gls{vn}'s traffic requirements, are proposed in Secs.~\ref{ssec:max_datarate} and \ref{sec:minReq}, respectively.

\subsection{System and Channel Models} 
\label{sub:system_model}
Let $\mathcal{M}$ be the set of \glspl{vn} and $\mathcal{N}$ be the set of \glspl{bs}.
In particular, $\mathcal{N}_m\subseteq \mathcal{N}$ is the set of BSs operating in the mmWave band, and $\mathcal{N}_L\subseteq \mathcal{N}$ is the set of BSs operating in the legacy band.
In \gls{v2i} networks, we expect other vehicles, pedestrians and environmental objects to block the link connecting the target \gls{vn} and its  serving \gls{bs}. It is therefore necessary to distinguish between \gls{los} and \gls{nlos} nodes.
For LTE, we consider the 3GPP model in~\cite{3gpp.36.842} for an outdoor  scenario. The \gls{los} path loss probability is given by 
\medmuskip=0mu
\thickmuskip=0mu
\eqPLLTE
\medmuskip=6mu
\thickmuskip=6mu
where $d$ is the distance in km between the \gls{vn} and the~candidate \gls{bs}, while the  complementary NLOS  probability is given by $\mathbb{P}_{\rm NLOS}^{\rm LTE}(d) = 1 - \mathbb{P}_{\rm LOS}^{\rm LTE}(d)$. 
For \gls{mmwave} cells, we consider the 3GPP model in~\cite{3gpp.38.901} for a UMi-Street-Canyon scenario. Accordingly, the \gls{los}  path loss probability is given~by 
\medmuskip=0mu
\thickmuskip=0mu
\eqPLmmw
\medmuskip=6mu
\thickmuskip=6mu
while its complementary NLOS probability is computed as $\mathbb{P}_{\rm NLOS}^{\rm mmW}(d) = 1 - \mathbb{P}_{\rm LOS}^{\rm mmW}(d)$. 
The path loss, for the \gls{los} and \gls{nlos} conditions, in dB, is given in~\cite{3gpp.36.842} and \cite{3gpp.38.901}, for LTE and mmWave respectively.

The channel quality between BS$_{j}$, $j\in\{1,\dots,|\mathcal{N}|\}$ and VN$_i$, $i\in\{1,\dots,|\mathcal{M}|\}$,  is measured in terms of SNR\footnote{Different PHY-layer metrics can be used to measure signal quality~\cite{3gpp.38.215}. In our paper, we chose to use the SNR (as considered in previous works, e.g., in \cite{giordani2018tutorial}).}, i.e.,
\begin{equation}\label{eq:snr}
 \text{SNR}_{ij} = 10\log_{10}\left(\frac{G_{ij} \cdot P_{tx}}{PL_{ij} \cdot N_0 B}\right),
\end{equation}
where $N_0$ is the thermal noise power spectral density, $B$ is the available bandwidth, $P_{tx}$ is the transmit power and $PL_{ij}$ is the path loss between BS$_{j}$  and VN$_i$. The term $G_{ij}$ represents the cumulative antenna gain between BS$_{j}$  and VN$_i$, which is a function of the number of antenna elements that each network node is equipped with. 
In case of LTE communications,  a single omnidirectional antenna is used, therefore $G_{ij}=1, \, \forall i, \, \forall j$. 
On the contrary, mmWave nodes form directional beams through Uniform Planar Arrays (UPAs) composed of multiple antenna elements, so that $G_{ij}\gg1$ when beam alignment is achieved~\cite{akdeniz2014millimeter}.

\subsection{Maximum SNR (MS) Policy}
\label{ssec:max_SNR}
The \gls{ms} policy represents one of the most common techniques for performing user association:  {VN$_i\in \mathcal{M}$} always connects to BS$_{j_{\rm MS}^*(i)}\in \mathcal{N}$ (either LTE or mmWave) that provides the maximum downlink average SNR, i.e.,  
\begin{equation}\label{eq:maxSNR}
j^*_{\rm MS}(i) = \argmax_{j\in\{1,\dots,|\mathcal{N}|\}} \{ \text{SNR}_{ij}\},\,\forall i\in \{1,\dots,|\mathcal{M}|\}
\end{equation}
where SNR$_{ij}$ is as in Eq.~\eqref{eq:snr}.
Notice that, in an urban heterogeneous scenario, the \gls{ms} policy does not guarantee that the \gls{bs} with the maximum \gls{snr} coincides with the closest one. 
First, LTE \glspl{bs} are generally preferred over mmWave ones due to the very low path loss experienced at below-6 GHz frequencies even at long distances.
Second, the mmWave signal is much more sensitive to penetration loss than LTE links and, therefore, if the geographically  closest \gls{mmwave} \gls{bs} is obstructed,  a further \gls{bs} in line of sight can potentially offer a better service (experiments performed for NLOS situations resulted in  SNR  degradation of more than 20 dB compared to LOS~propagation~\cite{rangan2014millimeter}).

We make the case that, although MS maximizes the SNR of vehicles, it does not  properly reflect the achievable end-to-end throughput of users, thereby leading to suboptimal association decisions. This is because, even with a lower SNR, mmWave cells may potentially deliver higher data rates (due to the much larger bandwidth) compared to LTE cells. 
Moreover, downlink-based received signal quality criteria do not characterize well uplink scenarios where vehicles have strict battery limitations on their transmit power.

\subsection{Maximum Rate (MR) Policy}
\label{ssec:max_datarate}
\gls{ms} attachment schemes can be improved by biasing cell selection with side information, e.g., network load.
A \gls{mr} approach is therefore proposed: {VN$_i\in \mathcal{M}$} connects to BS$_{j_{\rm MR}^*(i)}\in \mathcal{N}$ (either LTE or mmWave) that provides the maximum achievable data rate $R$,  i.e., 
\begin{equation}\label{eq:maxrate}
j^*_{\rm MR}(i) = \argmax_{j\in\{1,\dots,|\mathcal{N}|\}} \{ {R}_{ij}\}{},\,\forall i\in \{1,\dots,|\mathcal{M}|\}
\end{equation}

In this paper, the achievable data rate $R_{ij}$ between BS$_{j}$, $j\in\{1,\dots,|\mathcal{N}|\}$ and VN$_i$, $i\in\{1,\dots,|\mathcal{M}|\}$ is an indication of the cell's maximum capacity and  is computed from Shannon's formula as a function of the \gls{snr}, i.e., 
\begin{equation}\label{eq:datarate}
R_{ij} = \frac{B}{m_j}\log_2(1 + \text{SNR}_{ij})
\end{equation}
where $B$ is the available bandwidth and $m_j$ is the number of vehicles connected to BS$_j$. 
Our results therefore represent an upper bound for the throughput of the VNs, as we do not investigate the effect of medium access control mechanisms nor that of higher-layer retransmissions. 
We also assume that, if the measured \gls{snr} is below a predefined threshold SNR$_{\rm th}$, the data rate is equal to $0$.

The \gls{mr} strategy generally guarantees higher average throughput compared to the \gls{ms} approach~\cite{giordani2018efficient}.
However,  it is recognized that maximizing the data rate of all vehicles may result in an unfair data rate allocation~\cite{corroy2012dynamic}. 
In particular, the huge bandwidth available to \gls{mmwave} systems would make the load of mmWave cells much heavier than that of LTE ones, hence resulting in mmWave cells that are  congested. 

\subsection{Requirement-Aware (RA) Policy}
\label{sec:minReq}
To cope with MS and MR limitations, we propose a \gls{ra} attachment policy which simultaneously maintains fairness and balances the traffic load among the cells.
The association decision is therefore made as a function of the vehicle's individual \gls{qos} requirements and the availability of radio~resources.

In the context of \glspl{c-its}, we expect heterogeneous application requirements (e.g., in terms of throughput, latency, reliability) which, although not yet fully specified, have already been outlined by the 3GPP in~\cite{3gpp.22.186}.\footnote{In this work, four classes of vehicular traffic, with different throughput, latency and reliability constraints, are considered, as illustrated in Sec.~\ref{sec:simulation_scenario}.}
The \gls{ra} policy tries therefore to associate vehicles with strict reliability constraints (e.g., for advanced safety applications enabling semi- or fully-automated driving,  the required data rate is relatively low, although very high levels of reliability are expected due to the sensitive nature of the exchanged information) to LTE cells since the propagation characteristics of the  legacy spectrum generally deliver a good compromise between low end-to-end latency and high connection stability at long range.
On the contrary, \gls{mmwave} cells are selected to support those categories of applications with the boldest per user data rate requirements (e.g., extended sensor applications, which enhance a vehicle's perception range through dissemination of sensor observations)  but with looser reliability constraints. 
\glspl{vn} may therefore be able to exploit the whole available \gls{mmwave} bandwidth since less demanding \glspl{vn} are   associated to LTE cells.

Formalizing, VN$_i\in \mathcal{M}$  connects to BS$_{j_{\rm RA}^*(i)}\in \mathcal{N}$ (either LTE or mmWave) that satisfies the following conditions:
  \medmuskip=0mu
\thickmuskip=0mu
\begin{equation}\label{eq:minReq}
j_{\rm RA}^{*}(i) = 
  \begin{cases}
    j_{\rm MR|L}^{*}(i)  & \text{if } {R}_{ij^*_{\rm MR|L}(i)} > \bar{R}_{i}, \\ 
    j_{\rm MR}^{*}(i) & \text{otherwise},
  \end{cases}
  \,\forall i\in \{1,\dots,|\mathcal{M}|\}
\end{equation}
  \medmuskip=6mu
\thickmuskip=6mu
where $j_{\rm MR|L}^{*}(i) = \argmax_{j\in\{1,\dots,|\mathcal{N}_L|\}} \{ {R}_{ij}\}$ and $j_{\rm MR}^{*}(i)$ is as in Eq.~\eqref{eq:maxrate}.
In particular, the LTE BS offering maximum data rate is chosen if the offered data rate ${R}_{ij^*_{\rm MR|L}}$ is above the data rate $\bar{R}_{i}$ required by VN$_i$, otherwise the \gls{mr} policy is applied.

\begin{table}[t!]
\small
  \centering
  \vspace{0.23cm}
    \renewcommand{\arraystretch}{1}
  \begin{tabular}{@{}lll@{}}
  \toprule
    Parameter & Value & Description\\ \midrule
    $A$ & $1$ km$^2$ & Simulation area\\
    $h_{\rm BS}$ & $30$ m & Height of \gls{bs}\\
    $h_{\rm VN}$ & $2$ m & Height of \gls{vn}\\
    UPA$_{\rm BS}$ & $8\times8$ & BS antenna array\\
    UPA$_{\rm VN}$ & $4\times 4$ & VN antenna array\\
    $N_{\rm sim}$ & $2000$ & Simulation runs\\
    SNR$_{\rm th}$ & $-5$ dB & \gls{snr} threshold\\
  $f_{\rm L}$ & $2.4$ GHz & LTE central frequency \\
  $f_{\rm m}$ & $28$ GHz & mmWave central frequency \\
      $P_{\rm TX,L}$ & $46$ dBm & LTE BS TX power \\
    $P_{\rm TX,m}$ & $27$ dBm & mmWave BS TX power \\
    $B_{\rm L}$ & $20$ MHz &LTE bandwidth\\
    $B_{\rm m}$ & $1$ GHz &mmWave bandwidth\\
        $\lambda_{L}$ & $4$ BS/km$^2$ & LTE BS density \\
    $\lambda_{m}$ & $\{4,\dots80\}$ BS/km$^2$ & mmWave BS density \\
    \bottomrule
    \end{tabular}
  \caption{Simulation parameters.}
  \label{table:simParams}
\end{table}

\section{Performance Evaluation}
\label{sec:performance_evaluation}

In Sec.~\ref{sec:simulation_scenario}, we present the simulation scenario and parameters we consider in our analysis, in Sec.~\ref{sec:performance_metrics} we overview our performance metrics and in Sec.~\ref{sec:results} we compare the performance of the proposed V2I attachment mechanisms. 

\subsection{Simulation Scenario and Parameters}
\label{sec:simulation_scenario}

The parameters used in our simulations are based on realistic system design assumptions and are reported in Table~\ref{table:simParams}.
\paragraph{PHY Parameters} 
\label{par:phy_parameters}
For LTE systems, BSs operate at 2.4 GHz through omnidirectional transmissions and leverage 20 MHz of bandwidth. For mmWave systems, BSs operate at 28 GHz with 1 GHz of bandwidth and are equipped with UPAs of $8 \times 8$ elements to form directional beams.
The transmission power $P_{\rm TX}$ is set to 46 dBm and 27 dBm for LTE and mmWave BSs, respectively.\medskip

\begin{figure}[t!]
  \setlength\fwidth{0.42\textwidth}
  \setlength\fheight{0.5\textwidth}
  \vspace{-0.33cm}
%
%
\definecolor{mycolor1}{rgb}{0.00000,1.00000,1.00000}%

\pgfplotsset{
	tick label style={font=\scriptsize},
	label style={font=\scriptsize},
	legend  style={font=\scriptsize}
}

\begin{tikzpicture}

\begin{axis}[%
width=0.951\fwidth,
height=0.75\fwidth,
at={(0\fwidth,0\fwidth)},
scale only axis,
xmin=0,
xmax=1000,
xlabel style={font=\color{white!15!black}\scriptsize},
xlabel={X [m]},
ymin=0,
ymax=1000,
ylabel style={font=\color{white!15!black}\scriptsize},
ylabel={Y [m]},
axis background/.style={fill=white},
axis x line*=bottom,
axis y line*=left,
xmajorgrids,
ymajorgrids,
legend style={font=\scriptsize,legend cell align=left, align=left, draw=white!15!black, at={(0.5,1.0)},/tikz/every even column/.append style={column sep=0.35cm},
	anchor=north ,legend columns=-1},
/pgf/number format/.cd,
1000 sep={}
]
\draw[pattern=north west lines, pattern color=blue, draw=black] (axis cs:250,250) rectangle (axis cs:750,750);
\addplot[only marks, mark=o, mark options={}, mark size=2.0pt, draw=black] table[row sep=crcr]{%
x	y\\
175.728757258508	46.9250474915678\\
332.424760430253	250.259412989876\\
556.353812906086	596.279454412316\\
774.321794120845	782.083771401901\\
416.991601566316	855.577067639451\\
534.688738221125	789.735314533839\\
417.973121278014	195.55002530504\\
727.654844938909	403.190596915495\\
822.074560610987	238.864297311225\\
647.41365399137	507.783352264551\\
712.198717411948	125.03489523085\\
935.158575690728	328.556098457531\\
456.888796597808	745.761806976462\\
695.747358917973	217.409980258225\\
258.455862683852	703.339022378897\\
461.655794914926	499.400675215744\\
417.277968016792	403.627654093575\\
941.459939075077	900.947565174822\\
321.766597092514	612.233164547208\\
927.422118963092	621.45712943266\\
730.885541878913	946.663558385362\\
786.182780156642	644.726502169953\\
784.969545465407	971.806921173823\\
530.390071260005	907.593504474674\\
259.533644175882	880.307022106119\\
358.297099585248	503.397469635158\\
773.710245349855	362.73496188476\\
234.83159804976	768.002476616112\\
181.127914027588	274.957283507239\\
217.465294214968	300.598397570036\\
215.514780304436	377.904360294208\\
682.738747845449	445.675261586094\\
438.540805679217	172.994842378132\\
338.675370211205	273.226948614067\\
769.706092304381	292.314337805263\\
22.0539952751414	493.986661214546\\
33.8199269680778	256.692091305311\\
516.094314905729	532.555340222697\\
486.490192805806	16.1458800733223\\
954.487242063053	990.152445105358\\
796.966616385692	751.647304940877\\
674.799719622787	203.829537650339\\
147.16417082647	591.649183493138\\
918.14807181074	366.367796307537\\
734.124452857187	764.367423896267\\
630.796292885523	429.098173920308\\
820.58285659163	781.079044648267\\
116.366789855827	703.739645356017\\
824.458751905719	691.297499975653\\
511.509335744564	617.71335386276\\
697.741370188971	938.02206528271\\
723.140992597364	157.094293488221\\
27.5421989334074	758.576179246214\\
788.577568044064	880.290520291829\\
792.374620785023	666.039905118881\\
722.247352317914	537.509568656709\\
492.999450379626	656.650272344473\\
775.746834033679	81.7334643997574\\
738.364244190068	722.973528586229\\
661.332716084738	39.83062068728\\
522.823365020307	943.818839558359\\
481.828810834437	466.293322808285\\
654.394364886857	690.953756205649\\
522.939211298002	794.399920730403\\
378.883669409135	888.251338674014\\
362.172911496256	13.1070298566485\\
982.433397248547	672.691237973239\\
398.568469183521	723.236063345752\\
689.826814563155	918.628868701092\\
427.418047373824	936.533218035053\\
634.613760567002	213.595070167389\\
818.917325307137	774.261768915685\\
647.654258213215	826.818735572578\\
794.348323973872	403.006829378408\\
601.216463565447	23.3792965586201\\
646.829625254411	901.622224554302\\
627.238033089502	768.744427114866\\
461.468108938416	740.963359295123\\
227.199325415586	189.56322515333\\
975.359897314534	946.424297354849\\
};
\addlegendentry{\gls{mmwave} \gls{bs}}

\addplot[only marks, mark=triangle*, mark options={fill=black}, mark size=3.5pt, draw=black] table[row sep=crcr]{%
	x	y\\
	459.345594758644	418.6541000369\\
	632.905540790967	527.178832946562\\
	961.211140174115	789.010114556244\\
	496.695509122595	211.141000894238\\
};
\addlegendentry{\gls{lte} \gls{bs}}

\end{axis}

\begin{axis}[%
width=1.227\fwidth,
height=0.92\fwidth,
at={(-0.16\fwidth,-0.101\fwidth)},
scale only axis,
xmin=0,
xmax=1,
ymin=0,
ymax=1,
axis line style={draw=none},
ticks=none,
axis x line*=bottom,
axis y line*=left,
legend style={legend cell align=left, align=left, draw=white!15!black}
]
\end{axis}
\end{tikzpicture}%
  \caption{Example of simulation scenario in which LTE and mmWave BSs are deployed according to a PPP of density $\lambda_L=4$ BS/km$^2$ and $\lambda_m=80$  BS/km$^2$, respectively.}
  \label{fig:scenario}
\end{figure}
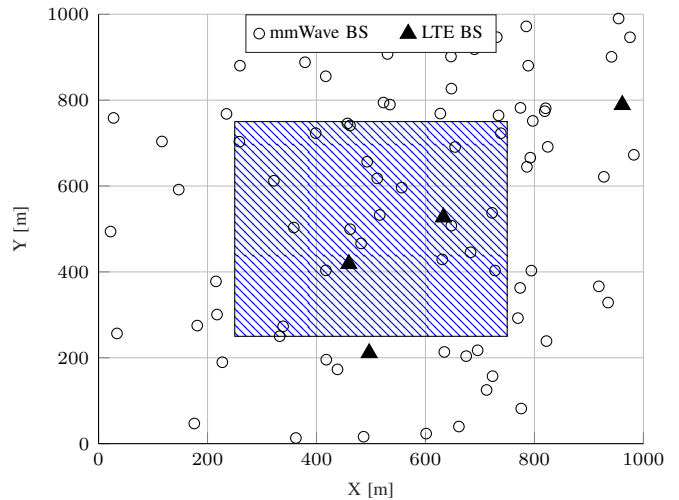

\paragraph{BS Deployment} 
\label{par:deployment}
BSs are deployed according to a \gls{ppp} of density $\lambda_L=4$ BS/km$^2$ for LTE BSs and  $\lambda_m$ spanning from $4$ to $80$ BS/km$^2$ for  mmWave BSs, over an area $A$ of $1$ km$^2$, as illustrated in \fig{fig:scenario}.\medskip

\paragraph{VN Deployment} 
VNs are uniformly  deployed over $A$ but, to avoid boundary effects, we collect statistics of just the VNs in a subset of the simulation area (the colored area in \fig{fig:scenario}).
We consider (i) a heavily loaded scenario in which an average of 10 vehicles per mmWave BS are deployed (so that the actual number of VNs in the network is a function of $\lambda_m$),  as foreseen in~\cite{3gpp.38.913}, or (ii) a scenario in which exactly $M=500$ VNs are deployed overall.
To evaluate the steady-state behavior of the network, \gls{vn}s' deployment consists of two steps,  following the approach used in \cite{Rebato17}. 
Said $\mathcal{M}$ the set of \glspl{vn} and $\mathcal{N}$ the set of \glspl{bs} (both \gls{lte} and \glspl{mmwave}), in the first step each \gls{vn}$_i \in \mathcal{M}$ is attached to \gls{bs}${_j^*} \in \mathcal{N}$ according to either of the algorithms described in \secref{sec:attachment}. 
Once all \glspl{vn} are attached to the network, in the second step  we iteratively update the cell association  by randomly picking one VN at a time. 
We repeat this  procedure by re-allocating a random VN at each step for a fairly large number of iterations, until convergence to the long-term VN distribution among the BSs is achieved. \medskip

\paragraph{Vehicular traffic classes} 
\label{par:vehicular_traffic_classes}

In the context of \glspl{c-its}, following the description in~\cite{3gpp.22.186}, we consider four different categories of vehicular traffic, to reflect the heterogeneity of future \gls{v2i} applications' characteristics and requirements. 
\begin{itemize}  
 \item \textit{Class 1 (e.g., basic safety)} Very high levels of communication stability are required, due to the potential consequences of communication errors, although data rates are below 1 Mbps.
 \item \textit{Class 2 (e.g., advanced safety, collision avoidance)} High degrees of reliability are required, and data rates of at least 10 Mbps should be supported.
 \item \textit{Class 3 (e.g., infotainment, cooperative perception)} Data rate requirements are in the order of hundreds of Mbps, while latency is reasonably tolerated.
 \item \textit{Class 4 (e.g., semi- or fully- automated driving through extended sensors)} The data rate demands are proportional to the resolution of the exchanged sensory data and likely exceed 1000 Mbps for high-quality uncompressed camera  measurements (e.g., ProRes 4444 with 4K resolution requires around 1200 Mbps). Latency requirements depend on the desired degree of automation.
\end{itemize}
Let $\mathcal{M}_k \subseteq \mathcal{M}$ be the subset of \glspl{vn} belonging to class $k$, $k\in\{1,\dots,4\}$. 
A \gls{vn} is assigned to one of these classes with equal probability,\footnote{A thorough analysis on the impact of $\mathbb{P}_k$ on the attachment performance is out of the scope of this paper and will be part of our future work.} so that
\begin{equation}
  \begin{split}
  \mathbb{P}_k &= \mathbb{P}[\text{VN} \in \mathcal{M}_k] = 0.25 \quad \forall k\in\{1,\dots,4\}
  \end{split}
\end{equation}

\begin{table}[t!]
\small
  \centering
  \vspace{0.23cm}
    \renewcommand{\arraystretch}{1}
  \begin{tabular}{@{}lll@{}}
  \toprule
    Traffic Class & Description & Data Rate\\ \midrule
    Class 1 & 	Basic safety & 1 Mbps \\
    Class 2 & 	Advanced safety & 10 Mbps \\
    Class 3 & 	Cooperative Perception & $>$ 100 Mbps \\
    Class 4 & 	Automated Driving & $>$ 1000 Mbps \\
    \bottomrule
    \end{tabular}
  \caption{Vehicular traffic classes requirements.}
  \label{table:classParams}
\end{table}

\subsection{Performance Metrics}
\label{sec:performance_metrics}
The performance evaluation is conducted as a function of the density of mmWave BSs and the attachment policy.
Our results are obtained following a Montecarlo method in which $N_{\rm sim}$ simulations are repeated  to make the conclusions statistically robust.
Let $R_{ij^*}$ be the data rate that VN$_i\in\mathcal{M}$ experiences when attached to the best BS$_{j^*(i)}\in\mathcal{N}$ according to either of the attachment policies described in Sec.~\ref{sec:attachment}. In particular, we consider the following performance metrics.

\begin{itemize}
  \item \emph{Mean data rate per class $\mathbb{E}[R]_k$}, which is computed as the sum of the data rates experienced by VNs belonging to traffic class $k$, divided by the total number of \glspl{vn} of that class, i.e., 
  \begin{equation}\label{eq:av_rate}
  \mathbb{E}[R]_k = \frac{\sum_{i \in \mathcal{M}_k} R_{ij^*(i)}}{|\mathcal{M}_k|} \quad \forall k\in\{1,\dots,4\}
  \end{equation}
  \item \emph{Mean data rate of the 10\% worst VNs per class $P_{{10}_k}$}, the average data rate relative to the worst 10\% of VNs of class $k$ (which represents the average performance of cell-edge nodes, i.e., the most resource-constrained network entities). 
  \item \emph{Percentage of \gls{vn}s satisfied $p_{{\rm sat}}$}, the percentage of VNs in the network that reach the minimum data rate requirements specified by the corresponding traffic class, as characterized in Sec.~\ref{sec:simulation_scenario}, i.e.,
  \medmuskip=3mu
\thickmuskip=3mu
\psat
\medmuskip=6mu
\thickmuskip=6mu
  where $\bar{R_i}$ is the data rate requirement for VN$_i\in\mathcal{M}$.
  \item \emph{Percentage of \gls{vn}s attached to \gls{lte} $p_{\rm LTE}$},  the percentage of \glspl{vn} in the network that are served by an LTE BS.
  \item \emph{Jain's fairness index $J_k$}, which gives an indication on whether network resources are shared fairly among the \glspl{vn}. This index is computed separately for each traffic class, and is defined as in~\cite{giordani2018efficient}, i.e.,
  \begin{equation}\label{eq:jains}
  J_k = \frac{\left(\sum_{i \in \mathcal{M}_k} R_{ij^*(i)}\right)^2}{|\mathcal{M}_k|\sum_{i \in \mathcal{M}_k} R^2_{ij^*(i)}} \quad \forall k\in\{1,\dots,4\}
  \end{equation}
  
\end{itemize}

\subsection{Results and Discussion}
\label{sec:results}

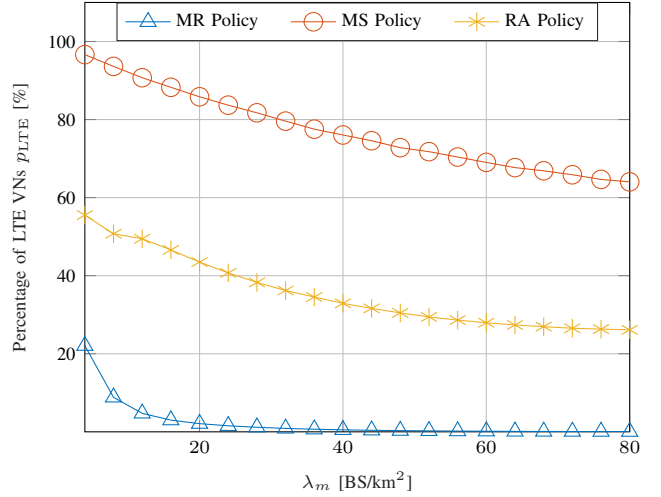
\begin{figure}[t!]
\centering
   \vspace{0.23cm}
  \setlength\fwidth{0.42\textwidth}
  \setlength\fheight{0.4\textwidth}
%
%
\definecolor{mycolor1}{rgb}{0.00000,0.44700,0.74100}%
\definecolor{mycolor2}{rgb}{0.85000,0.32500,0.09800}%
\definecolor{mycolor3}{rgb}{0.92900,0.69400,0.12500}%

\pgfplotsset{
	tick label style={font=\scriptsize},
	label style={font=\scriptsize},
	legend  style={font=\scriptsize}
}
\begin{tikzpicture}

\begin{axis}[%
width=0.951\fwidth,
height=0.75\fwidth,
at={(0\fwidth,0\fwidth)},
scale only axis,
xmin=4,
xmax=80,
xlabel style={font=\color{white!15!black}\scriptsize},
xlabel={$\lambda_m$ [BS/km$^2$]},
ymin=0,
ymax=1.1,
ytick={0.2, 0.4,0.6,0.8,1},
yticklabels={20, 40, 60, 80,100},
ylabel style={font=\color{white!15!black}\scriptsize},
ylabel={Percentage of LTE VNs $p_{\rm LTE}$ [\%]},
axis background/.style={fill=white},
xmajorgrids,
ymajorgrids,
legend style={font=\scriptsize,legend cell align=left, align=left, draw=white!15!black, at={(0.5,1)},/tikz/every even column/.append style={column sep=0.35cm},
	anchor=north ,legend columns=-1},
]
\addplot [color=mycolor1, mark size=3.5pt, mark=triangle, mark options={solid, mycolor1}]
  table[row sep=crcr]{%
4	0.220707317073171\\
8	0.0883827160493827\\
12	0.0473429752066116\\
16	0.0299378881987578\\
20	0.020952736318408\\
24	0.0154253112033195\\
28	0.0116459074733096\\
32	0.00874454828660436\\
36	0.00665927977839335\\
40	0.00508229426433915\\
44	0.00393990929705215\\
48	0.00311954261954262\\
52	0.00244433781190019\\
56	0.00190819964349376\\
60	0.00149417637271215\\
64	0.00110764430577223\\
68	0.000949339207048458\\
72	0.000789875173370319\\
76	0.000624835742444152\\
80	0.000483770287141074\\
};
\addlegendentry{MR Policy}

\addplot [color=mycolor2, mark size=3.5pt, mark=o, mark options={solid, mycolor2}]
  table[row sep=crcr]{%
4	0.966378048780488\\
8	0.936148148148148\\
12	0.907194214876033\\
16	0.882627329192547\\
20	0.858333333333333\\
24	0.836877593360996\\
28	0.817284697508897\\
32	0.79596261682243\\
36	0.775385041551247\\
40	0.760462593516209\\
44	0.745777777777778\\
48	0.727974012474012\\
52	0.717724568138196\\
56	0.7039688057041\\
60	0.690280366056572\\
64	0.676869734789391\\
68	0.668730543318649\\
72	0.658575589459085\\
76	0.646711563731932\\
80	0.640310237203496\\
};
\addlegendentry{MS Policy}

\addplot [color=mycolor3, mark size=3.5pt, mark=asterisk, mark options={solid, mycolor3}]
  table[row sep=crcr]{%
4	0.55519512195122\\
8	0.507395061728395\\
12	0.494528925619835\\
16	0.465996894409938\\
20	0.434898009950249\\
24	0.407118257261411\\
28	0.382749110320285\\
32	0.361971962616822\\
36	0.345387811634349\\
40	0.329170822942643\\
44	0.316221088435374\\
48	0.30472869022869\\
52	0.294381957773512\\
56	0.286167557932264\\
60	0.279490848585691\\
64	0.273690327613105\\
68	0.269375917767988\\
72	0.265312760055478\\
76	0.263168199737188\\
80	0.261245942571785\\
};
\addlegendentry{RA Policy}

\end{axis}
\end{tikzpicture}%
  \caption{Percentage $p_{\rm LTE}$ of VNs attached to the \gls{lte} BSs vs. the mmWave BS density $\lambda_m$, for different attachment policies.  A heavily loaded scenario in which an average of 10 vehicles per mmWave BS are deployed is considered.}
  \label{fig:ratio_users}
\end{figure}

\textbf{LTE Associations. } In~\fig{fig:ratio_users} we plot the percentage of \glspl{vn} served by \gls{lte} BSs, which gives an overview of how vehicles are distributed across the network.
As expected, the different propagation characteristics of sub- and above-6GHz bands and the high imbalance in the available network resources could result in different conclusions as a function of  $\lambda_m$ and the attachment policy.
In general, the \gls{ms} approach tends to associate most \glspl{vn} to \gls{lte} \glspl{bs} ($p_{\rm LTE} \geq 60\%$ for all investigated density configurations) since they transmit with a higher power, have a larger communication range and are less affected by propagation and absorption loss compared to mmWave BSs. 
On the contrary, \glspl{vn} are generally attached to mmWave BSs if the \gls{mr} approach is preferred, since mmWave systems offer opportunities for order of magnitude higher data rates than operating at LTE, even at low SNR.
Notice that, for the \gls{mr} case, $p_{\rm LTE}$ decreases for increasing values of $\lambda_m$ as a consequence of stronger mmWave channels (in case of sparsely deployed networks, i.e., $\lambda_m<20$ BS/km$^2$, many mmWave BSs are in outage, thereby making LTE cells a desirable attachment solution despite the limited available bandwidth).
\fig{fig:ratio_users} also demonstrates that the \gls{ra} policy guarantees more fair resource utilization with respect to its counterparts.
VNs are indeed almost equally distributed across LTE and mmWave BSs: class 1 and 2 VNs (i.e., around $50\%$ of the overall traffic) are associated to LTE BSs, the only network entities satisfying strict reliability constraints, while class 3 and 4 VNs are associated to mmWave BSs to satisfy bold data rate requirements. 
For high values of $\lambda_m$, $p_{\rm LTE}$ finally converges to $25\%$, that corresponds to the percentage of VNs belonging to class 1.

\begin{figure}[t!]
\centering
  \vspace{0.23cm}
  \setlength\fwidth{0.42\textwidth}
  \setlength\fheight{0.4\textwidth}
%
%
\definecolor{mycolor1}{rgb}{0.00000,0.44700,0.74100}%
\definecolor{mycolor2}{rgb}{0.85000,0.32500,0.09800}%
\definecolor{mycolor3}{rgb}{0.92900,0.69400,0.12500}%

\pgfplotsset{
	tick label style={font=\scriptsize},
	label style={font=\scriptsize},
	legend  style={font=\scriptsize}
}

\begin{tikzpicture}

\begin{axis}[%
width=0.951\fwidth,
height=0.74\fwidth,
at={(0\fwidth,0\fwidth)},
scale only axis,
xmin=4,
xmax=80,
xlabel style={font=\color{white!15!black}\scriptsize},
xlabel={$\lambda_m$ [BS/km$^2$]},
ymin=0,
ymax=1400,
ylabel style={font=\color{white!15!black}\scriptsize},
ylabel={Mean data rate  $\mathbb{E}[R]_1$ [Mbps]},
axis background/.style={fill=white},
xmajorgrids,
ymajorgrids,
legend style={font=\scriptsize,legend cell align=left, align=left, draw=white!15!black, at={(0.5,1.0)},/tikz/every even column/.append style={column sep=0.1cm},
	anchor=north ,legend columns=-1},
	/pgf/number format/.cd,
1000 sep={}
]
\addplot [color=mycolor1, mark size=3.5pt, mark=triangle, mark options={solid, mycolor1}]
  table[row sep=crcr]{%
4	539.369466216339\\
8	662.028664243413\\
12	765.568825040116\\
16	831.815958795851\\
20	899.015564445523\\
24	953.379020756972\\
28	998.645206245307\\
32	1028.13296039984\\
36	1061.54963635528\\
40	1084.72048826619\\
44	1114.56405868579\\
48	1137.4284612767\\
52	1157.82001774464\\
56	1175.74336987063\\
60	1193.70968072721\\
64	1211.56023016681\\
68	1224.6512322157\\
72	1240.99884663005\\
76	1249.37146797289\\
80	1260.31952626753\\
};
\addlegendentry{MR Policy}

\addplot [color=mycolor2, mark size=3.5pt, mark=o, mark options={solid, mycolor2}]
  table[row sep=crcr]{%
4	245.923693734969\\
8	435.899324402117\\
12	567.529428034653\\
16	597.592945406663\\
20	699.15242907795\\
24	752.143042821769\\
28	804.282717456802\\
32	834.286944383996\\
36	882.036722124179\\
40	879.603158815387\\
44	904.42598648592\\
48	917.776117425963\\
52	959.283386652488\\
56	968.804611820136\\
60	985.495584284491\\
64	1003.25451464783\\
68	992.940072201189\\
72	1015.87200024983\\
76	1025.73621286782\\
80	1035.52925018425\\
};
\addlegendentry{MS Policy}

\addplot [color=mycolor3, mark size=3.5pt, mark=asterisk, mark options={solid, mycolor3}]
  table[row sep=crcr]{%
4	54.2592113436971\\
8	30.0023728527101\\
12	20.5205813913311\\
16	16.4066449913028\\
20	14.0027685723652\\
24	12.5181662625539\\
28	11.418037705683\\
32	10.608430210204\\
36	9.8773061551288\\
40	9.358419063101\\
44	8.84353964083411\\
48	8.42168369587155\\
52	7.99710438284335\\
56	7.71402462195259\\
60	7.30385606028874\\
64	6.99982201173527\\
68	6.71358538754894\\
72	6.45964935647216\\
76	6.16693503369359\\
80	5.88156306502968\\
};
\addlegendentry{RA Policy}

 \addplot[mark=none, blue,line width=3.7, dashed] coordinates {(0,1) (80,1)};
  \addlegendentry{Requirement}

\coordinate (pt) at (axis cs:75,40);
\end{axis}

\node[pin={[pin distance=0.5cm]93:{%
	\begin{tikzpicture}[trim axis left,trim axis right]
	\begin{axis}[
	axis background/.style={fill=white},
	tiny,
	xmin=70,xmax=80,
	ymin=0,ymax=10,
	enlargelimits,
	xmajorgrids,
	ymajorgrids,
	]
	\addplot [color=mycolor3, mark size=3.5pt, mark=asterisk, mark options={solid, mycolor3}]
	table[row sep=crcr]{%
		4	54.2592113436971\\
		8	30.0023728527101\\
		12	20.5205813913311\\
		16	16.4066449913028\\
		20	14.0027685723652\\
		24	12.5181662625539\\
		28	11.418037705683\\
		32	10.608430210204\\
		36	9.8773061551288\\
		40	9.358419063101\\
		44	8.84353964083411\\
		48	8.42168369587155\\
		52	7.99710438284335\\
		56	7.71402462195259\\
		60	7.30385606028874\\
		64	6.99982201173527\\
		68	6.71358538754894\\
		72	6.45964935647216\\
		76	6.16693503369359\\
		80	5.88156306502968\\
	};

	 \addplot[mark=none, dashed, blue,line width=3.75] coordinates {(0,1) (90,1)};
	\end{axis}
	\end{tikzpicture}%
}}] at (pt) {};
\draw [draw=black] (6.1,0) rectangle (7, 0.3);

\end{tikzpicture}%
  \caption{Mean data rate $\mathbb{E}[R]_1$ for VNs of class 1 vs. $\lambda_m$, for different attachment policies. A heavily loaded scenario in which an average of 10 vehicles per mmWave BS are deployed is considered. }
  \label{fig:ratePerClass1}
\end{figure}
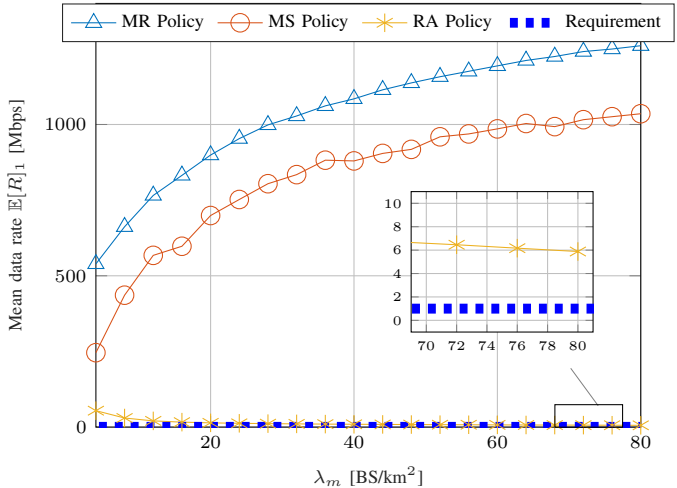

\begin{figure}[t]
\centering
  \setlength\fwidth{0.41\textwidth}
  \setlength\fheight{0.4\textwidth}
%
%
\definecolor{mycolor1}{rgb}{0.00000,0.44700,0.74100}%
\definecolor{mycolor2}{rgb}{0.85000,0.32500,0.09800}%
\definecolor{mycolor3}{rgb}{0.92900,0.69400,0.12500}%

\pgfplotsset{
	tick label style={font=\scriptsize},
	label style={font=\scriptsize},
	legend  style={font=\scriptsize}
}

\begin{tikzpicture}

\begin{axis}[%
width=0.951\fwidth,
height=0.75\fwidth,
at={(0\fwidth,0\fwidth)},
scale only axis,
xmin=4,
xmax=80,
xlabel style={font=\color{white!15!black}\scriptsize},
xlabel={$\lambda_m$ [BS/km$^2$]},
ymin=0,
ymax=1800,
ylabel style={font=\color{white!15!black}\scriptsize},
ylabel={Mean data rate  $\mathbb{E}[R]_4$ [Mbps]},
axis background/.style={fill=white},
xmajorgrids,
ymajorgrids,
legend style={font=\scriptsize,legend cell align=left, align=left, draw=white!15!black, at={(0.5,1.10)},/tikz/every even column/.append style={column sep=0.1cm},
	anchor=north ,legend columns=-1},
		/pgf/number format/.cd,
1000 sep={}
]
\addplot [color=mycolor1, mark size=3.5pt, mark=triangle, mark options={solid, mycolor1}]
  table[row sep=crcr]{%
4	535.16753409361\\
8	664.343626645417\\
12	759.15524271644\\
16	834.237886490728\\
20	902.896873405435\\
24	954.265368261059\\
28	999.004283786659\\
32	1026.26204088131\\
36	1061.50423906331\\
40	1089.10102300209\\
44	1114.33407762444\\
48	1134.5755643088\\
52	1158.0496457547\\
56	1177.72020890834\\
60	1192.83518937336\\
64	1212.05957485605\\
68	1224.86027255677\\
72	1239.2888169023\\
76	1249.44023148028\\
80	1260.67220203236\\
};
\addlegendentry{MR Policy}

\addplot [color=mycolor2,  mark size=3.5pt, mark=o, mark options={solid, mycolor2}]
  table[row sep=crcr]{%
4	284.523974929098\\
8	426.499312450554\\
12	525.137362424072\\
16	663.577249144477\\
20	663.040886539354\\
24	749.4573989403\\
28	811.076178025529\\
32	819.407988612431\\
36	864.424116267549\\
40	862.973884105722\\
44	921.054076552967\\
48	925.582112204112\\
52	948.45478178548\\
56	952.745066768975\\
60	985.925060513017\\
64	1000.61694115255\\
68	1002.54983659773\\
72	1014.01168025465\\
76	1016.52046086932\\
80	1033.79470695372\\
};
\addlegendentry{MS Policy}

\addplot [color=mycolor3,  mark size=3.5pt, mark=asterisk, mark options={solid, mycolor3}]
  table[row sep=crcr]{%
4	938.568222884846\\
8	1224.49799860745\\
12	1440.97579225484\\
16	1525.43729486742\\
20	1572.78905862353\\
24	1584.39585843876\\
28	1600.75628781314\\
32	1604.77242454989\\
36	1609.23380315643\\
40	1617.72847271239\\
44	1622.51444389834\\
48	1626.62871426737\\
52	1633.11387066753\\
56	1635.37405318749\\
60	1640.8622378838\\
64	1648.16800955237\\
68	1664.30003272089\\
72	1664.82318432619\\
76	1684.26103265147\\
80	1690.68155280752\\
};
\addlegendentry{RA Policy}

 \addplot[mark=none, blue,line width=3.7, dashed] coordinates {(0,1200) (80,1200)};
  \addlegendentry{Requirement}

\end{axis}

\begin{axis}[%
width=1.227\fwidth,
height=0.82\fwidth,
at={(-0.16\fwidth,-0.101\fwidth)},
scale only axis,
xmin=0,
xmax=1,
ymin=0,
ymax=1,
axis line style={draw=none},
ticks=none,
axis x line*=bottom,
axis y line*=left,
legend style={legend cell align=left, align=left, draw=white!15!black}
]
\end{axis}
\end{tikzpicture}%
  \caption{Mean data rate $\mathbb{E}[R]_4$ for VNs of class 4 vs.  $\lambda_m$, for different attachment policies. A heavily loaded scenario in which an average of 10 vehicles per mmWave BS are deployed is considered. }
  \label{fig:ratePerClass4}
\end{figure}
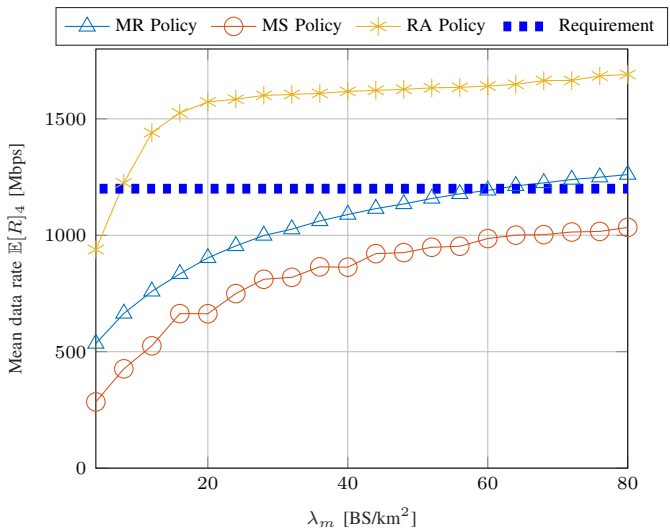

\textbf{Data Rate.} As another performance measure, in Figs. \ref{fig:ratePerClass1} and \ref{fig:ratePerClass4} we plot the average data rate that class 1 and class 4 VNs experience, respectively, when implementing either of the attachment policies presented in Sec.~\ref{sec:attachment}.
We consider a heavily loaded scenario in which an average of 10 vehicles per mmWave BS are deployed.
At first glance we observe that, while all investigated attachment schemes satisfy class~1 data rate requirements, i.e.,  1 Mbps (Fig.~\ref{fig:ratePerClass1}),  MS and MR are generally not able to sustain class 4  requests, i.e.,  1200 Mbps for 4K resolution cameras (Fig.~\ref{fig:ratePerClass4}), thereby making the proposed RA solution the only viable approach to  maximize the communication performance for all categories of vehicular services.
MR eventually meets class 4 requirements, though only for very high values of $\lambda_m$ (i.e., $\lambda_m>70$ BS/km$^2$); such ultra-dense deployment, however, could be costly for network operators, in terms of capital and management expenditures, and should therefore be avoided.
With the MS approach, VNs connect to BSs showing the instantaneous highest signal strengths and avoid instead nodes that provide lower SNR values (but possibly higher rates, due to their low traffic loads).
With the MR approach, VNs connect to mmWave BSs and share the same amount of radio resources regardless of the individual traffic requirements, with class 1 VNs experiencing much higher date rate than requested, at the expense of class 4 VNs experiencing poor date rate in overloaded cells.
On the other hand, the RA strategy, which biases association decisions with side information about vehicle requirements, tends to associate class 1 VNs to LTE cells  (which, despite the limited capacity of the physical channel, can easily support class 1's rate requests) and saves network bandwidth for those categories of VNs with the most stringent connectivity demands.
Numerically, for class 4 VNs, RA delivers up to 1.5 times higher throughput compared to MR and a  2 fold throughput increase compared to~MS.

\begin{figure}[t]
\centering
  \vspace{0.23cm}
  \setlength\fwidth{0.41\textwidth}
  \setlength\fheight{0.4\textwidth}
%
%
\definecolor{mycolor1}{rgb}{0.00000,0.44700,0.74100}%
\definecolor{mycolor2}{rgb}{0.85000,0.32500,0.09800}%
\definecolor{mycolor3}{rgb}{0.92900,0.69400,0.12500}%
\pgfplotsset{
	tick label style={font=\scriptsize},
	label style={font=\scriptsize},
	legend  style={font=\scriptsize}
}

\begin{tikzpicture}

\begin{axis}[%
width=0.951\fwidth,
height=0.74\fwidth,
at={(0\fwidth,0\fwidth)},
scale only axis,
xmin=4,
xmax=80,
xlabel style={font=\color{white!15!black}\scriptsize},
xlabel={$\lambda_m$ [BS/km$^2$]},
ymin=0,
ymax=3000,
ytick={500, 1000,1500,2000,2500},
yticklabels={500, 1000,1500,2000,2500},
ylabel style={font=\color{white!15!black}\scriptsize},
ylabel={Mean data rate $\mathbb{E}[R]_4$ [Mbps]},
axis background/.style={fill=white},
xmajorgrids,
ymajorgrids,
legend style={font=\scriptsize,legend cell align=left, align=left, draw=white!15!black, at={(0.5,1.0)},/tikz/every even column/.append style={column sep=0.1cm},
	anchor=north ,legend columns=-1},
		/pgf/number format/.cd,
1000 sep={}
]
\addplot [color=mycolor1,  mark size=3.5pt, mark=triangle, mark options={solid, mycolor1}]
  table[row sep=crcr]{%
4	45.0520184609826\\
8	109.739319691631\\
12	187.195489297253\\
16	273.59870223732\\
20	364.957985500636\\
24	460.879266704274\\
28	559.80052833499\\
32	663.792771029522\\
36	769.055428258985\\
40	876.148902739428\\
44	985.300927705944\\
48	1092.09636527894\\
52	1207.27292353902\\
56	1317.80080163486\\
60	1432.34617589074\\
64	1544.462782443\\
68	1666.56519276898\\
72	1776.95377290154\\
76	1894.25760278089\\
80	2011.31630186421\\
};
\addlegendentry{MR Policy}

\addplot [color=mycolor2,  mark size=3.5pt, mark=o, mark options={solid, mycolor2}]
  table[row sep=crcr]{%
4	73.6974231934488\\
8	142.509912260093\\
12	221.211235286859\\
16	296.943534997361\\
20	370.747217461713\\
24	447.849810256169\\
28	520.819701725937\\
32	613.472538202516\\
36	682.479747835656\\
40	745.4906361796\\
44	830.551954221165\\
48	897.270677711006\\
52	966.817562003547\\
56	1077.5327743104\\
60	1124.38196021103\\
64	1182.88329811506\\
68	1256.67425462096\\
72	1338.4318808337\\
76	1408.50787903635\\
80	1500.75970043952\\
};
\addlegendentry{MS Policy}

\addplot [color=mycolor3,  mark size=3.5pt, mark=asterisk, mark options={solid, mycolor3}]
  table[row sep=crcr]{%
4	54.35646623909\\
8	142.101126591249\\
12	252.456234157914\\
16	376.803937652483\\
20	507.32557697055\\
24	648.593929703857\\
28	789.441854154725\\
32	934.350133859092\\
36	1090.52800863453\\
40	1238.44935334167\\
44	1396.65072741991\\
48	1546.39389503029\\
52	1705.75192252835\\
56	1878.04757727805\\
60	2029.45827817336\\
64	2196.72437044014\\
68	2356.6835303308\\
72	2519.07958323701\\
76	2692.82627367731\\
80	2852.58139391956\\
};
\addlegendentry{RA Policy}

\addplot+[color=red,fill=red!10!white,const plot, dashed, opacity=0.7, line width=0.5, mark=none,forget plot]
    coordinates {(0,3000) (40,3000)}\closedcycle;
    \node[color=black, font=\scriptsize] at (22,2600) {Densely loaded network};
    \addplot+[color=green,fill=green!10!white,const plot, dashed, opacity=0.7, line width=0.5, mark=none,forget plot]
    coordinates {(40,3000) (80,3000)}\closedcycle;
    \node[color=black, font=\scriptsize] at (60,2600) {Scarcely loaded network};
 \addplot[mark=none, blue,line width=3.7, dashed] coordinates {(0,1200) (80,1200)};
  \addlegendentry{Requirement}

\end{axis}

\begin{axis}[%
width=1.227\fwidth,
height=0.79\fwidth,
at={(-0.16\fwidth,-0.1\fwidth)},
scale only axis,
xmin=0,
xmax=1,
ymin=0,
ymax=1,
axis line style={draw=none},
ticks=none,
axis x line*=bottom,
axis y line*=left,
legend style={legend cell align=left, align=left, draw=white!15!black}
]
\end{axis}
\end{tikzpicture}%
  \caption{Mean data rate $\mathbb{E}[R]_4$ for VNs of class 4 vs.  $\lambda_m$, for different attachment policies. A scenario in which 500 VNs are deployed overall is~considered. }
  \label{fig:500ratePerClass4}
\end{figure}
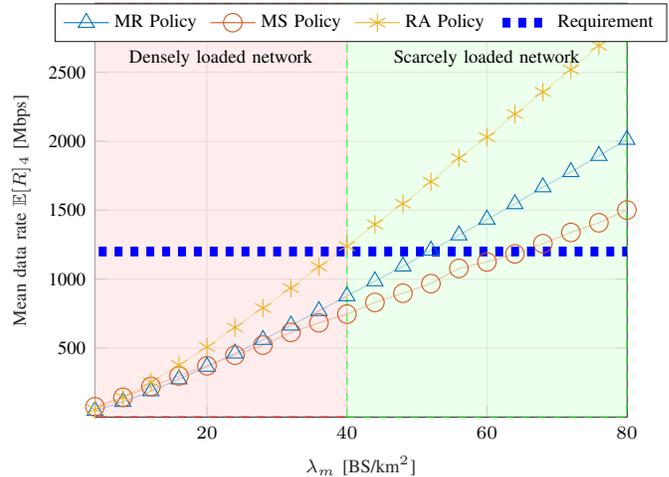

The same conclusions can be drawn from \fig{fig:500ratePerClass4}, which compares the attachment performance of different network loading regimes, considering a scenario in which 500 class 4 VNs are deployed overall.
We see that, for densely loaded scenarios (i.e., $\lambda_m<40$ BS/km$^2$), none of the investigated attachment policies satisfies the requirement of $1200$ Mbps. 
Conversely, the throughput linearly increases with $\lambda_m$ since each BS serves fewer vehicles and can handle traffic requests more efficiently. Moreover, densification guarantees that the endpoints are progressively closer, thus guaranteeing improved signal quality and higher received power.
The effect of densification is particularly evident for the RA cases (e.g., the data rate increases  more than 5 times from 20 to 80 BS/km$^2$).
Moreover, \fig{fig:500ratePerClass4} demonstrates that the RA policy guarantees higher average throughput than any other attachment strategy (e.g., $+40\%$  compared to the MR approach for $\lambda_m=40$ BS/km$^2$).
The performance gap is even more significant when the density $\lambda_m$ is increased, thereby moving from NLOS to LOS propagation.

\begin{figure}[t]
\centering
  \setlength\fwidth{0.41\textwidth}
  \setlength\fheight{0.4\textwidth}
%
%
\definecolor{mycolor1}{rgb}{0.00000,0.44700,0.74100}%
\definecolor{mycolor2}{rgb}{0.85000,0.32500,0.09800}%
\definecolor{mycolor3}{rgb}{0.92900,0.69400,0.12500}%

\pgfplotsset{
	tick label style={font=\scriptsize},
	label style={font=\scriptsize},
	legend  style={font=\scriptsize}
}

\begin{tikzpicture}

\begin{axis}[%
width=0.951\fwidth,
height=0.7\fwidth,
at={(0\fwidth,0\fwidth)},
scale only axis,
xmin=4,
xmax=80,
xlabel style={font=\color{white!15!black}\scriptsize},
xlabel={$\lambda_m$ [BS/km$^2$]},
ymin=0,
ymax=1250,
ylabel style={font=\color{white!15!black}\scriptsize},
ylabel={Mean data rate of worst VNs $P_{{10}_4}$ [Mbps]},
axis background/.style={fill=white},
xmajorgrids,
ymajorgrids,
legend style={font=\scriptsize,legend cell align=left, align=left, draw=white!15!black, at={(0.5,1.0)},/tikz/every even column/.append style={column sep=0.35cm},
	anchor=north ,legend columns=-1},
/pgf/number format/.cd,
1000 sep={}
]
\addplot [color=mycolor1, mark=triangle, mark size=3.5pt,mark options={solid, mycolor1}]
  table[row sep=crcr]{%
4	313.953684491023\\
8	319.113764192272\\
12	359.698154029672\\
16	403.366767107319\\
20	464.470789056438\\
24	504.856223035641\\
28	541.641107830975\\
32	572.757077737424\\
36	602.64274271107\\
40	632.862987165633\\
44	659.849903785886\\
48	683.070419753366\\
52	704.216937782326\\
56	729.564747623634\\
60	746.322120503999\\
64	768.883167445571\\
68	785.432884660199\\
72	797.222342078309\\
76	813.266490263203\\
80	827.966127639556\\
};
\addlegendentry{MR Policy}

\addplot [color=mycolor2, mark=o, mark size=3.5pt,mark options={solid, mycolor2}]
  table[row sep=crcr]{%
4	89.5905480366794\\
8	18.564484809808\\
12	7.65218428206134\\
16	4.93061892435069\\
20	4.04749408917897\\
24	3.48769435707254\\
28	3.11909319330482\\
32	2.80111573103787\\
36	2.56985921224114\\
40	2.37205041593518\\
44	2.23703211672376\\
48	2.10068533938846\\
52	2.00932200196596\\
56	1.91156341237471\\
60	1.83493000207186\\
64	1.76317750307266\\
68	1.71390719314103\\
72	1.64318425588959\\
76	1.72820841982938\\
80	1.55936628241695\\
};
\addlegendentry{MS Policy}

\addplot [color=mycolor3, mark=asterisk, mark size=3.5pt,mark options={solid, mycolor3}]
  table[row sep=crcr]{%
4	542.064044105301\\
8	578.111930215614\\
12	652.653957427204\\
16	722.138100277966\\
20	780.478927036448\\
24	817.155857593945\\
28	848.881345677844\\
32	880.549941868427\\
36	897.527778712587\\
40	925.913752499765\\
44	945.925761197294\\
48	966.882552962025\\
52	984.770484306143\\
56	996.927001207342\\
60	1013.84588827042\\
64	1028.9481304729\\
68	1050.14294418315\\
72	1059.08092246707\\
76	1078.13270210396\\
80	1091.75152573\\
};
\addlegendentry{RA Policy}

\end{axis}

\begin{axis}[%
width=1.227\fwidth,
height=0.85\fwidth,
at={(-0.16\fwidth,-0.1\fwidth)},
scale only axis,
xmin=0,
xmax=1,
ymin=0,
ymax=1,
axis line style={draw=none},
ticks=none,
axis x line*=bottom,
axis y line*=left,
legend style={legend cell align=left, align=left, draw=white!15!black}
]
\end{axis}
\end{tikzpicture}%
  \caption{Mean data rate for the 10\% worst VNs of class 4 vs.  $\lambda_m$, for different attachment policies. A heavily loaded scenario in which an average of 10 vehicles per mmWave BS are deployed is considered.}
  \label{fig:10perc}
\end{figure}
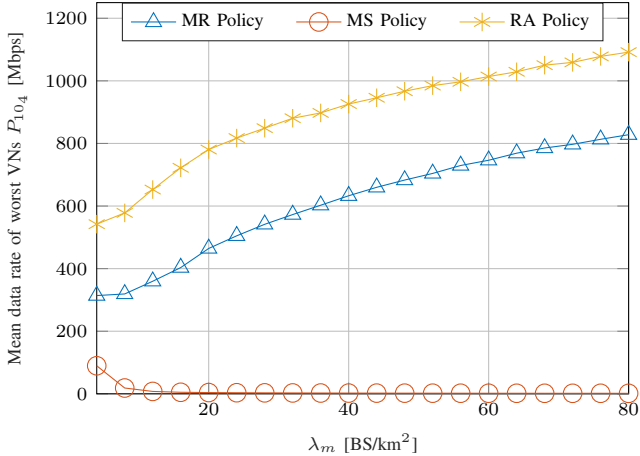

The patterns we observed in the previous plots can be recognized considering also the data rate relative to the worst 10\% of the users, i.e., cell-edge VNs, as illustrated in  Fig.~\ref{fig:10perc}.
Class 4 VNs are considered.
We see that cell-edge VNs experience a significant data rate decrease with respect to the average values measured in Fig.~\ref{fig:ratePerClass4}, motivating efforts towards network densification to increase the coverage of cell-edge vehicles.
Moreover, although none of the investigated attachment policies can meet class 4's requirements, the RA approach still outperforms both MR and MS strategies in terms of data rate ($+65\%$ and an impressive $+4600\%$, respectively,  for $\lambda_m=40$ BS/km$^2$).

\textbf{Fairness.} Although fairness is not necessarily a pre-requisite for V2I systems (e.g., safety-critical operations shall deserve prioritization), it still represents a major concern that should be taken into account to guarantee a minimum level of performance to the cell-edge users (or, in general, to users experiencing bad channel conditions).
In Fig.~\ref{fig:jains1} we plot  Jain's fairness index $J_1$ for class 1 VNs. 
We demonstrate that the RA solution, which associates \gls{vn}s with low data rate requirements to LTE cells, guarantees  more fair cell association compared to traditional attachment schemes. 
On one side, MS strategies homogeneously attach VNs to \gls{lte} when a few BSs are deployed but, as $\lambda_{m}$ gets higher, start associating some of the VNs to \gls{mmwave} BSs too (see Fig.~\ref{fig:ratio_users}), thereby offering completely different access channels.
On the other side, MR strategies  attach VNs to \gls{mmwave} cells which are generally not compatible with fairness as a result of the increased variability of the above-6 GHz channel.
However,  for high values of $\lambda_m$, i.e., when  pushing the network into LOS regimes, MR's fairness performance is deemed comparable to that of RA.

\begin{figure}[t]
\centering
  \setlength\fwidth{0.42\textwidth}
  \setlength\fheight{0.4\textwidth}
%
%
\definecolor{mycolor1}{rgb}{0.00000,0.44700,0.74100}%
\definecolor{mycolor2}{rgb}{0.85000,0.32500,0.09800}%
\definecolor{mycolor3}{rgb}{0.92900,0.69400,0.12500}%

\pgfplotsset{
	tick label style={font=\scriptsize},
	label style={font=\scriptsize},
	legend  style={font=\scriptsize}
}

\begin{tikzpicture}

\begin{axis}[%
width=0.951\fwidth,
height=0.7\fwidth,
at={(0\fwidth,0\fwidth)},
scale only axis,
xmin=4,
xmax=80,
xlabel style={font=\color{white!15!black}\scriptsize},
xlabel={$\lambda_m$ [BS/km$^2$]},
ymin=0,
ymax=1.1,
ylabel style={font=\color{white!15!black}\scriptsize},
ylabel={Jain's Fairness Index J$_1$},
axis background/.style={fill=white},
xmajorgrids,
ymajorgrids,
legend style={font=\scriptsize,legend cell align=left, align=left, draw=white!15!black, at={(0.5,1.0)},/tikz/every even column/.append style={column sep=0.35cm},
	anchor=north ,legend columns=-1},
]
\addplot [color=mycolor1, mark size=3.5pt, mark=triangle, mark options={solid, mycolor1}]
  table[row sep=crcr]{%
4	0.840049381212785\\
8	0.828485456807007\\
12	0.845779953332018\\
16	0.869071683299215\\
20	0.886353249948015\\
24	0.899098747343968\\
28	0.908122283895382\\
32	0.916519466152866\\
36	0.923194951702375\\
40	0.928509282798462\\
44	0.932895150522186\\
48	0.937143722693005\\
52	0.940678551089658\\
56	0.943362139962377\\
60	0.946187089663251\\
64	0.947713887159976\\
68	0.950223383186487\\
72	0.951666399501006\\
76	0.953718156639365\\
80	0.954756069057191\\
};
\addlegendentry{MR Policy}

\addplot [color=mycolor2, mark size=3.5pt, mark=o, mark options={solid, mycolor2}]
  table[row sep=crcr]{%
4	0.891762727575084\\
8	0.765763026185201\\
12	0.61891875382327\\
16	0.512352975079861\\
20	0.408633431806878\\
24	0.342278044152019\\
28	0.293343975910231\\
32	0.255656430030429\\
36	0.225640683732848\\
40	0.202297988888953\\
44	0.189566099495877\\
48	0.183708004534154\\
52	0.182630818426421\\
56	0.182876215258013\\
60	0.189183452837221\\
64	0.196703961361189\\
68	0.188921191964859\\
72	0.194465000309376\\
76	0.197943322485943\\
80	0.201681050634711\\
};
\addlegendentry{MS Policy}

\addplot [color=mycolor3, mark size=3.5pt, mark=asterisk, mark options={solid, mycolor3}]
  table[row sep=crcr]{%
4	0.973705757081089\\
8	0.959966820069605\\
12	0.956162588014919\\
16	0.953953589053953\\
20	0.952283635680443\\
24	0.952770250582592\\
28	0.95153917244934\\
32	0.950871279571723\\
36	0.951320602563982\\
40	0.950970577854741\\
44	0.949816624995789\\
48	0.950787334083241\\
52	0.949874598682577\\
56	0.949878491430034\\
60	0.949317603419023\\
64	0.948739679155949\\
68	0.948617827540011\\
72	0.948475902234756\\
76	0.948076251552183\\
80	0.948268602044193\\
};
\addlegendentry{RA Policy}

\end{axis}

\begin{axis}[%
width=1.227\fwidth,
height=0.85\fwidth,
at={(-0.16\fwidth,-0.1\fwidth)},
scale only axis,
xmin=0,
xmax=1,
ymin=0,
ymax=1,
axis line style={draw=none},
ticks=none,
axis x line*=bottom,
axis y line*=left,
legend style={legend cell align=left, align=left, draw=white!15!black}
]
\end{axis}
\end{tikzpicture}%
  \caption{Jain's  index $J_1$ for \glspl{vn} of class 1 vs.  $\lambda_m$, for different attachment policies. A heavily loaded scenario in which an average of 10 vehicles per mmWave BS are deployed is considered.}
  \label{fig:jains1}
\end{figure}

\begin{figure}[t]
\centering
  \setlength\fwidth{0.42\textwidth}
  \setlength\fheight{0.4\textwidth}
%
%
\definecolor{mycolor1}{rgb}{0.00000,0.44700,0.74100}%
\definecolor{mycolor2}{rgb}{0.85000,0.32500,0.09800}%
\definecolor{mycolor3}{rgb}{0.92900,0.69400,0.12500}%

\pgfplotsset{
	tick label style={font=\scriptsize},
	label style={font=\scriptsize},
	legend  style={font=\scriptsize}
}

\begin{tikzpicture}

\begin{axis}[%
width=0.951\fwidth,
height=0.7\fwidth,
at={(0\fwidth,0\fwidth)},
scale only axis,
xmin=4,
xmax=80,
xlabel style={font=\color{white!15!black}\scriptsize},
xlabel={$\lambda_m$ [BS/km$^2$]},
ymin=0.3,
ymax=1,
ytick={0.2, 0.4,0.6,0.8,1},
yticklabels={20, 40, 60, 80,100},
ylabel style={font=\color{white!15!black}\scriptsize},
ylabel={Percentage of users satisfied $p_{\rm SAT}$ [\%]},
axis background/.style={fill=white},
xmajorgrids,
ymajorgrids,
legend style={font=\scriptsize,legend cell align=left, align=left, draw=white!15!black, at={(0.5,1.10)},/tikz/every even column/.append style={column sep=0.35cm},
	anchor=north ,legend columns=-1},
]
\addplot [color=mycolor1, mark size=3.5pt, mark=triangle, mark options={solid, mycolor1}]
  table[row sep=crcr]{%
4	0.775801461698441\\
8	0.797148397516477\\
12	0.811671755871339\\
16	0.827368804980758\\
20	0.842551651906342\\
24	0.855155918691783\\
28	0.86720130580046\\
32	0.875846793388636\\
36	0.887537291055247\\
40	0.895969029017185\\
44	0.904201153457724\\
48	0.912208322200692\\
52	0.919423074740598\\
56	0.925846720442094\\
60	0.930951512847849\\
64	0.937410684681273\\
68	0.941458610574437\\
72	0.94556911764205\\
76	0.949038266923088\\
80	0.953041735502486\\
};
\addlegendentry{MR Policy}

\addplot [color=mycolor2, mark size=3.5pt, mark=o, mark options={solid, mycolor2}]
  table[row sep=crcr]{%
4	0.519143403438238\\
8	0.457364285538487\\
12	0.400948412540114\\
16	0.364575122430241\\
20	0.357042522693332\\
24	0.359288737721517\\
28	0.366367518730088\\
32	0.373476393655027\\
36	0.384815350660912\\
40	0.387686353139447\\
44	0.403844652477654\\
48	0.408958403417029\\
52	0.417068291705326\\
56	0.42415421854281\\
60	0.43254613374005\\
64	0.444051228560003\\
68	0.443297481549855\\
72	0.453506393379493\\
76	0.458565286817776\\
80	0.463632449262319\\
};
\addlegendentry{MS Policy}

\addplot [color=mycolor3, mark size=3.5pt, mark=asterisk, mark options={solid, mycolor3}]
  table[row sep=crcr]{%
4	0.825489083076798\\
8	0.880147716850076\\
12	0.917819208842713\\
16	0.935537718136208\\
20	0.948587588589403\\
24	0.956244634364858\\
28	0.9624772417748\\
32	0.966837726052061\\
36	0.969896365100761\\
40	0.973766946915063\\
44	0.975848496483493\\
48	0.977982749581538\\
52	0.979984447898687\\
56	0.980683319036851\\
60	0.982057457650005\\
64	0.982280417830339\\
68	0.984172480987866\\
72	0.983956681714025\\
76	0.984893511273483\\
80	0.985200741055356\\
};
\addlegendentry{RA Policy}

\end{axis}

\begin{axis}[%
width=1.227\fwidth,
height=0.85\fwidth,
at={(-0.16\fwidth,-0.101\fwidth)},
scale only axis,
xmin=0,
xmax=1,
ymin=0,
ymax=1,
axis line style={draw=none},
ticks=none,
axis x line*=bottom,
axis y line*=left,
legend style={legend cell align=left, align=left, draw=white!15!black}
]
\end{axis}
\end{tikzpicture}%
  \caption{Percentage $p_{\rm sat}$ of VSs satisfied vs. the mmWave BS density $\lambda_m$, for different attachment policies. A heavily loaded scenario in which an average of 10 vehicles per mmWave BS are deployed is considered.}
  \label{fig:satisfied}
\end{figure}

\textbf{Percentage of VNs Satisfied.}
Finally, it is interesting to compare the three attachment algorithms in terms of percentage of VNs which satisfy application demands. 
We see that the MS approach, which tries to associate vehicles to LTE cells, is penalized by class 4 VNs whose very rigid data rate requirements cannot be sustained by low-bandwidth LTE connections. 
The performance particularly degrades when $\lambda_m\geq5$ BS/km$^2$, i.e., when the number of VNs in the network starts increasing as a result of denser mmWave deployments, and then slightly increases when $\lambda_m\geq20$ BS/km$^2$, i.e., when VNs that connect to mmWaves find LOS BSs.
On the other hand, we observe that, although the MR approach guarantees a good level of satisfaction among the vehicles, i.e., $p_{\rm sat} > 85\%$ for highly dense networks, with the RA scheme more than $95\%$ of VNs are able to meet QoS demands even in low-density deployments.
This is because RA discriminates association requests as a function of QoS requirements and balances VNs between LTE and mmWave BSs avoiding the overload of transmission links.
Based on the above discussion, we therefore make the case that the proposed framework represents the most appropriate attachment strategy to maximize the communication~performance.

\section{Concluding Remarks}
\label{sec:conclusions}

In this work we proposed a novel requirement-aware attachment strategy that delivers fair, robust and efficient vehicle association in heterogeneous networks in which both \gls{mmwave} and \gls{lte} cellular infrastructures are deployed.
In particular, we show that benchmark methods which bias  cell selection decisions with received  signal quality or network load information  cannot support those categories of vehicular traffic with the boldest connectivity requirements, and can therefore lead to sub-optimal association.
On the contrary, we demonstrated that the proposed approach, which makes attachment decisions as a function of the vehicular service requirements, prevents the overload of transmission links and represents the most appropriate strategy to meet QoS demands even considering low-density deployments.

As part of our future work, we will extend our implementation including advanced offloading techniques that distribute vehicles among the network cells even after the initial attachment decisions.
Moreover, we will validate our simulation framework with an accurate mathematical analysis based on  stochastic geometry or combinatorial optimization.

\bibliography{biblio}

\begin{thebibliography}{10}

\bibitem{bengler2014three}
K.~Bengler, K.~Dietmayer, B.~Farber, M.~Maurer, C.~Stiller, and H.~Winner,
  ``{Three decades of driver assistance systems: Review and future
  perspectives},'' {\em IEEE Intelligent Transportation Systems Magazine},
  vol.~6, pp.~6--22, Oct 2014.

\bibitem{araniti2013lte}
G.~Araniti, C.~Campolo, M.~Condoluci, A.~Iera, and A.~Molinaro, ``{LTE for
  vehicular networking: a survey},'' {\em IEEE Communications Magazine},
  vol.~51, pp.~148--157, May 2013.

\bibitem{3gpp.22.186}
3GPP, ``{Enhancement of 3GPP support for V2X scenarios; Stage 1},'' Technical
  Specification (TS) 22.186, 2018.

\bibitem{lu2014connected}
N.~Lu, N.~Cheng, N.~Zhang, X.~Shen, and J.~W. Mark, ``{Connected vehicles:
  Solutions and challenges},'' {\em IEEE Internet of Things Journal}, vol.~1,
  pp.~289--299, May 2014.

\bibitem{choi2016millimeter}
J.~Choi, V.~Va, N.~Gonzalez-Prelcic, R.~Daniels, C.~R. Bhat, and R.~W. Heath,
  ``{Millimeter-wave vehicular communication to support massive automotive
  sensing},'' {\em IEEE Communications Magazine}, vol.~54, pp.~160--167, Dec
  2016.

\bibitem{giordani2017millimeter}
M.~Giordani, A.~Zanella, and M.~Zorzi, ``{Millimeter wave communication in
  vehicular networks: Challenges and opportunities},'' in {\em 6th
  International Conference on Modern Circuits and Systems Technologies
  (MOCAST)}, 2017.

\bibitem{zheng2015heterogeneous}
K.~Zheng, Q.~Zheng, P.~Chatzimisios, W.~Xiang, and Y.~Zhou, ``{Heterogeneous
  vehicular networking: A survey on architecture, challenges, and solutions},''
  {\em IEEE Communications Surveys \& Tutorials}, vol.~17, pp.~2377--2396, Jun
  2015.

\bibitem{Dhillon13}
H.~S. Dhillon, R.~K. Ganti, and J.~G. Andrews, ``Load-aware modeling and
  analysis of heterogeneous cellular networks,'' {\em IEEE Transactions on
  Wireless Communications}, vol.~12, pp.~1666--1677, April 2013.

\bibitem{chen2011joint}
C.~S. Chen, F.~Baccelli, and L.~Roullet, ``{Joint optimization of radio
  resources in small and macro cell networks},'' in {\em IEEE 73rd Vehicular
  Technology Conference (VTC Spring)}, 2011.

\bibitem{corroy2012dynamic}
S.~Corroy, L.~Falconetti, and R.~Mathar, ``{Dynamic cell association for
  downlink sum rate maximization in multi-cell heterogeneous networks},'' in
  {\em IEEE International Conference on Communications (ICC)}, pp.~2457--2461,
  2012.

\bibitem{jo2012heterogeneous}
H.-S. Jo, Y.~J. Sang, P.~Xia, and J.~G. Andrews, ``{Heterogeneous cellular
  networks with flexible cell association: A comprehensive downlink SINR
  analysis},'' {\em IEEE Transactions on Wireless Communications}, vol.~11,
  pp.~3484--3495, Aug 2012.

\bibitem{lasaulce2011game}
S.~Lasaulce and H.~Tembine, {\em Game theory and learning for wireless
  networks: fundamentals and applications}.
\newblock Academic Press, 2011.

\bibitem{papadimitriou1998combinatorial}
C.~H. Papadimitriou and K.~Steiglitz, {\em Combinatorial optimization:
  algorithms and complexity}.
\newblock Courier Corporation, 1998.

\bibitem{bethanabhotla2014user}
D.~Bethanabhotla, O.~Y. Bursalioglu, H.~C. Papadopoulos, and G.~Caire, ``User
  association and load balancing for cellular massive mimo,'' in {\em
  Information Theory and Applications Workshop (ITA)}, 2014.

\bibitem{xu2017user}
Y.~Xu and S.~Mao, ``{User association in massive MIMO HetNets},'' {\em IEEE
  Systems Journal}, vol.~11, pp.~7--19, Sep 2017.

\bibitem{Liu14}
D.~Liu, Y.~Chen, K.~K. Chai, T.~Zhang, and M.~Elkashlan, ``{Opportunistic User
  Association for Multi-Service HetNets Using Nash Bargaining Solution},'' {\em
  IEEE Comm. Letters}, vol.~18, pp.~463--466, March 2014.

\bibitem{Liu16}
H.~Liu, Z.~Gao, X.~Shao, and W.~Zhou, ``{A centralized user association scheme
  for load balancing and UE energy efficiency in HetNets},'' in {\em 2nd IEEE
  International Conference on Computer and Communications (ICCC)},
  pp.~2965--2969, Oct 2016.

\bibitem{ye2013user}
Q.~Ye, B.~Rong, Y.~Chen, M.~Al-Shalash, C.~Caramanis, and J.~G. Andrews, ``User
  association for load balancing in heterogeneous cellular networks,'' {\em
  IEEE Transactions on Wireless Communications}, vol.~12, pp.~2706--2716, Apr
  2013.

\bibitem{baccelli2010stochastic}
F.~Baccelli, B.~B{\l}aszczyszyn, {\em et~al.}, ``{Stochastic geometry and
  wireless networks: Volume I Theory},'' {\em Foundations and
  Trends{\textregistered} in Networking}, vol.~3, no.~34, pp.~249--449, 2010.

\bibitem{elsawy2013stochastic}
H.~ElSawy, E.~Hossain, and M.~Haenggi, ``Stochastic geometry for modeling,
  analysis, and design of multi-tier and cognitive cellular wireless networks:
  A survey,'' {\em IEEE Communications Surveys \& Tutorials}, vol.~15,
  pp.~996--1019, Jun 2013.

\bibitem{Cheung12}
W.~C. Cheung, T.~Q.~S. Quek, and M.~Kountouris, ``Throughput optimization,
  spectrum allocation, and access control in two-tier femtocell networks,''
  {\em IEEE Journal on Selected Areas in Communications}, vol.~30,
  pp.~561--574, April 2012.

\bibitem{Singh2015}
S.~Singh, M.~N. Kulkarni, A.~Ghosh, and J.~G. Andrews, ``{Tractable Model for
  Rate in Self-Backhauled Millimeter Wave Cellular Networks},'' {\em IEEE
  Journal on Selected Areas in Communications}, vol.~33, pp.~2196--2211, May
  2015.

\bibitem{giordani2018efficient}
M.~Giordani, M.~Mezzavilla, S.~Rangan, and M.~Zorzi, ``{An Efficient Uplink
  Multi-Connectivity Scheme for 5G Millimeter-Wave Control Plane
  Applications},'' {\em IEEE Transactions on Wireless Communications}, vol.~17,
  pp.~6806--6821, Aug 2018.

\bibitem{liang2018towards}
L.~{Liang}, H.~{Ye}, and G.~Y. {Li}, ``Toward intelligent vehicular networks: A
  machine learning framework,'' {\em IEEE Internet of Things Journal}, vol.~6,
  pp.~124--135, Feb 2019.

\bibitem{li2017user}
Z.~Li, C.~Wang, and C.~Jiang, ``{User Association for Load Balancing in
  Vehicular Networks: An Online Reinforcement Learning Approach},'' {\em IEEE
  Transactions on Intelligent Transportation Systems}, vol.~18, pp.~2217--2228,
  Aug 2017.

\bibitem{giordani2016initial}
M.~Giordani, M.~Mezzavilla, and M.~Zorzi, ``{Initial access in 5G mmWave
  cellular networks},'' {\em IEEE Communications Magazine}, vol.~54,
  pp.~40--47, Nov 2016.

\bibitem{3gpp.36.842}
3GPP, ``{Study on Small Cell enhancements for E-UTRA and E-UTRAN; Higher layer
  aspects},'' Technical Report (TR) 36.842, 2014.

\bibitem{3gpp.38.901}
3GPP, ``{Study on channel model for frequencies from 0.5 to 100 GHz (Release
  14)},'' Technical Report (TR) 38.901, 2018.

\bibitem{3gpp.38.215}
3GPP, ``{NR—Physical Layer Measurements (Release 15)},'' Technical
  Specification (TS) 38.215, 2018.

\bibitem{giordani2018tutorial}
M.~Giordani, M.~Polese, A.~Roy, D.~Castor, and M.~Zorzi, ``{A Tutorial on Beam
  Management for 3GPP NR at mmWave Frequencies},'' {\em IEEE Communications
  Surveys \& Tutorials}, First Quarter 2019.

\bibitem{akdeniz2014millimeter}
M.~R. Akdeniz, Y.~Liu, M.~K. Samimi, S.~Sun, S.~Rangan, T.~S. Rappaport, and
  E.~Erkip, ``Millimeter wave channel modeling and cellular capacity
  evaluation,'' {\em IEEE Journal on Selected Areas in Communications},
  vol.~32, pp.~1164--1179, June 2014.

\bibitem{rangan2014millimeter}
S.~Rangan, T.~S. Rappaport, and E.~Erkip, ``{Millimeter-wave cellular wireless
  networks: Potentials and challenges},'' {\em Proceedings of the IEEE},
  vol.~102, pp.~366--385, Feb 2014.

\bibitem{3gpp.38.913}
3GPP, ``{Study on scenarios and requirements for next generation access
  technologies (Release 14)},'' Technical Report (TR) 38.913, 2018.

\bibitem{Rebato17}
M.~Rebato, F.~Boccardi, M.~Mezzavilla, S.~Rangan, and M.~Zorzi, ``Hybrid
  spectrum sharing in mmwave cellular networks,'' {\em IEEE Transactions on
  Cognitive Communications and Networking}, vol.~3, pp.~155--168, June 2017.

\end{thebibliography}
\bibliographystyle{ieeetr}

\end{document}